\documentclass[showkeys,reprint,
 bibnotes,
 amsmath,amssymb,
 aps,
floatfix,]{revtex4-1}

\usepackage{graphicx}
\usepackage{dcolumn}
\usepackage{bm}
\usepackage{bbm}
 \usepackage{booktabs}

\begin{document}


\title{An empirical evaluation of alternative methods of estimation for Permutation Entropy in time series with tied values}


\author{Francisco Traversaro }
\affiliation{ITBA}

\author{Marcelo Risk}
\affiliation{CONICET-ITBA}

\author{Osvaldo Rosso }
\affiliation{CONICET-HIBA}

\author{Francisco O. Redelico }
\affiliation{CONICET-HIBA}


\date{\today}

\begin{abstract}
 Bandt and Pompe introduced Permutation Entropy in 2002 for Time Series where equal values, $x_t^*=x_t, t^*=t$, were neglected and only inequalities between the $x_t$ were considered. Since then, this measure has been modified and extended, in particular in cases when the amount of equal values in the series can not be neglected, (i.e heart rate variability (HRV) time series). We review the different existing methodologies that treats this subject by classifying them according to their different strategies.
In addition, a novel Bayesian Missing Data Imputation is presented that proves to outperform the existing methodologies that deals with type of time series.   
All this facts are illustrated by simulations and also by distinguishing patients suffering from Congestive Heart Failure from a (healthy) control group using HRV time series. 
\end{abstract}


\maketitle

\section{Introduction \label{Introduction}}
In the seminal article of Bandt and Pompe \cite{bandt2002permutation}, when introducing Permutation Entropy it is stated as a condition for the estimation that the data of the time series is continuous, hence the probability of two equal values is equal to zero. 
In the rare event that ties exists, the suggestion is whether to ignore the patterns with ties, or to add small random perturbation. Unfortunately, that condition does not hold for many time series. Discrete time series with repeated values may occur by two reasons. Either the time series comes from a discrete data generator process, (e.g. a Poisson autoregressive model where the marginal probability distribution follows a Poisson member family \cite{mckenzie1988some}), or when the data of the generator process is continuous but only a coarse version of the actual realization is observed as a consequence of lack of precision of the data measuring device. The latter is the case of the heart rate frecuency (HRV) time series. This contribution deals with the changes necessary to analyse both kind of data, but emphasizes in the case when the Bandt and Pompe hypothesis of continuous data holds despite the deficiencies in the observation process. Several papers suggest modification in the estimation of Permutation Entropy to cope with repeated values. They either do this extending the symbolic alphabet presented by Bandt and Pompe, \cite{bian2012modified} or giving new rules to sort data, mostly to analyse HRV time series \cite{parlitz2012classifying}.
 We propose a new strategy to cope with this issue. This methodology uses the information of the actual time series to deal with patterns with ties. It assumes that this patterns are actually coming from suitable versions of the original patterns without ties and so they will contribute to the count of these patterns using an a priori probability distribution. It can be seen as a Bayesian Methodology. 
 
This contribution has a twofold objective. The first is to introduce a new method to deal with tied values and the second is to do an exhaustive exploration of all the methods presented. The paper reads as follows, Section \ref{OP} presents the two different existing mappings from real values to symbols, Section \ref{PEalg} reviews the different methodologies in the literature for calculating the Permutation Entropy of time series with repeated values and presents a novel and improving methodology, Section \ref{results} reviews the performance, in terms of the Mean Square Error (MSE) and Bias, of these strategies, Section \ref{realAP} shows a real application of classifying ECG signal, and finally Section \ref{Conclusions} is devoted to conclusions. 
\section{Ordinal Patterns and the mappings to a symbolic alphabet}\label{OP}

Let $\{X_t\}_{t \in T}$ be a realization of a data generator process in form of a real valued time series of length $T$, at first assuming $P(x_{t_1}=x_{t_2})=0~\forall t_1 \neq t_2$ (i.e there is not equal values in the time series). If the $\{X_t\}_{t \in T}$ attain infinitely many values, it is common to replace them by a symbol sequence $\{(\pi_i)_{t}\}$ with finitely many symbols, and calculate source entropy for the $\{(\pi_i)_{t}\}$ \cite{bandt2002permutation}.
Let $X_m(t)=(x_t,x_{t+1},\dots,x_{t+m-1})$ with $0\leq t \leq T-m+1$ be the embedded vectors of length $m$ of the time series $\{X_t\}_{t \in T}$.
Let $S_{m \geq 2}$ the symmetric group of order $m!$ form by all possible permutation of order $m$, $\pi_i=(i_1,i_2,\dots,i_m) \in~S_m$ ($i_j \neq i_k \forall j \neq k$ so every element in $\pi_i$ is unique). We will call an element $\pi_i$ in $S_m$ a symbol.
Then $X_m(t)$ can be mapped to a symbol $\pi_i$ in $S_m$. This mapping should be defined in a way that preserves the desired relation  between the elements $x_t$ in $X_m(t)$; and all $t \in T$ that shares this pattern has to mapped to the same element of $\pi_i \in S_m$. 

In the literature that encompass Permutation Entropy there are two ways to define the mapping between $X_m(i)=(x_i,x_{i+1},\dots,x_{i+m-1})$ to $\pi_i=(i_1,i_2,\dots,i_m) \in S_m$ (i.e mapping patterns onto symbols):
\begin{enumerate}
\item \textbf{Permutating the ranks}:  ordering the ranks (eq. \ref{eq:ranks}) of the $x_i$ in $X_m(t)$ in chronological order (i.e. \textit{Rank Permutation}) \cite{autocorrelationBandt,Riedl2013,Bandt2005}.
\item\textbf{Permutating the time indexes}: ordering the time indexes according to the ranks of $x_i$ in $X_m(t)$ (i.e. \textit{Chronological Index Permutation})\cite{bandt2002permutation,parlitz2012classifying,bian2012modified}.
\end{enumerate}

\subsection{Rank Permutation Mapping}\label{RankBA}
For a given but otherwise arbitrary $t$, the $m$ number of real values $X_m(t)=(x_t,x_{t+1},\dots,x_{t+m-1})$ are mapped onto their rank. The rank function is defined as: 
\begin{equation} \label{eq:ranks}
R(x_{t+n}) = \sum_{k=0}^{m-1} \mathbbm{1}(x_{t+k} \leq x_{t+n})
\end{equation}
where $ \mathbbm{1}$ is the indicator function (i.e $ \mathbbm{1}(Z) = 1$ if $Z$ is true and $0$ otherwise) , $x_{t+n} \in X_m(t)$ and $1 \leq R(x_{t+n}) \leq m$. So the rank $R(min(x_{t+k}))=1$ and $R(max(x_{t+k}))=m$. The complete alphabet is all the possible permutation of the ranks.  

Hence, any vector $X_m(t)$ is uniquely mapped onto $\pi_i=\left(R(x_t),R(x_{t+1}),\dots,R(x_{t+m-1})\right) \in S_m$. This means that each value $x_t$ in $X_m(t)$ is replaced for its rank.\\
For example, let us take the series with seven values \citep{bandt2002permutation}, and embedding dimension $m=3$:
\begin{equation}
X_t=(4,7,9,10,6,11,3) ~~T=7
\end{equation}

$X_3(1)=(4,7,9)$ and $X_3(2)=(7,9,10)$ represents the permutation $\pi = 123$ since $R(x_1)= 1$ ,$R(x_2)= 2$, $R(x_3)= 3$. $X_3(3)=(9,10,6)$ and $X_3(4)=(6,11,3)$ correspond to the permutation $\pi = 231$ since $R(x_1)= 2$ ,$R(x_2)= 3$, $R(x_3)= 1$.

With this Rank Rank Permutation Mapping one simply maps each value $x_i$ in $X_m(t)$ placing its rank $R(x_i) \in \{1,2,\dots,m\}$ in chronological order in $\pi_i$ in $S_m$. 
In Figure \ref{fig:rank} an illustrative drawing of this mapping for all alternatives in $m=3$ is presented. It can be seen that the indexes of the vertical axis are fixed, ordered by amplitude (i.e ranks), and they are mapped onto the time axis. The resultant symbol can be obtained reading the labels in the horizontal axis from left to right (in chronological order).
This method is used by \cite{autocorrelationBandt,Riedl2013,Bandt2005} among others.

\subsection{Chronological Index Permutation Mapping} 

Again, for a given but otherwise arbitrary $t$, the $m$ number of real values $X_m(t)=(x_t,x_{t+1},\dots,x_{t+m-1})$ can be rearranged in increasing order respect to their amplitude. In order to do the mapping to $\pi_i=(i_1,i_2,\dots,i_m) \in S_m$, $(i_1,i_2,\dots,i_m)$ must comply that $x_{t+\bm{i_1}-1} < x_{t+\bm{i_2}-1} < \dots <x_{t+\bm{i_m}-1}$. Thus, the time indexes are ordered according to their amplitude. The complete alphabet is all the possible permutation of these chronological indexes.

Lets take the previous series as an example:

$X_3(1)=(4,7,9)$ and $X_3(2)=(7,9,10)$  represents the permutation $\pi = 123$ since $x_{t+\textbf{1}} < x_{t+\textbf{2}}< x_{t+\textbf{3}}$. $X_3(3)=(9,10,6)$ and $X_3(4)=(6,11,3)$ correspond to the permutation $\pi = 312$ since $x_{t+\textbf{3}} < x_{t+\textbf{1}} < x_{t+\textbf{2}}$.

\begin{figure}[h!]
\includegraphics[width=80mm]{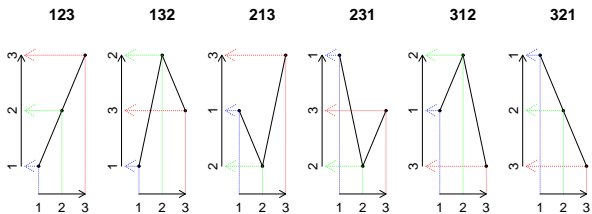}
\caption{\textbf{Chronological Index Permutation Mapping}: All symbols for $m=3$ are shown. This order based alphabet simply maps each value $x_i$ in $X_m(t)$ ordering its index $t \in \{1,2,\dots,m\}$ according to the increasing amplitude (rank) of each $x_i$ in $X_m(t)$. It can be seen that the indexes of the time axis are fixed in chronological order, and they are mapped onto the vertical (amplitude) axis. For each pattern $X_3(t)=(x_t,x_{t+1},x_{t+2})$, the resultant symbol $\pi_i \in S_3$ can be obtained reading the labels in the vertical axis from the bottom to the top (in the direction of the increasing amplitude).}
\label{fig:order}
\end{figure}

With this Chronological Index Permutation Mapping one simply maps each value $x_i$ in $X_m(t)$ ordering its time index $t \in \{1,2,\dots,m\}$ according to the increasing amplitude (rank) of each $x_i$ in $X_m(t)$. 
In Figure \ref{fig:order} an illustrative drawing of this mapping for all alternatives in $m=3$ is presented. It can be seen that the indexes of the time axis are fixed in chronological order, and they are mapped onto the vertical (amplitude) axis. The resultant symbol can be obtained reading the labels in the vertical axis from bottom to top (in the direction of the increasing amplitude). 

This method is used by \cite{bandt2002permutation,parlitz2012classifying,bian2012modified} among others.\\

\begin{figure}[h!]
\includegraphics[width=80mm]{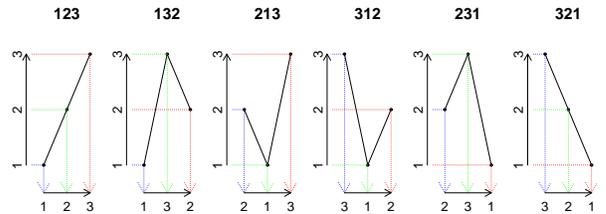}
\caption{\textbf{Rank Permutation Mapping} All symbols for $m=3$ are shown. With this Rank Alphabet one simply maps each value $x_i$ in $X_m(t)$ placing its rank $R(x_i) \in \{1,2,\dots,m\}$ in chronological order in $\pi_i$ in $S_m$.It can be seen that the indexes of the vertical axis are fixed, ordered by amplitude (i.e ranks), and they are mapped onto the time axis. For each pattern $X_3(t)=(x_t,x_{t+1},x_{t+2})$, the resultant symbol can be obtained reading the labels in the horizontal axis from left to right (in chronological order). }
\label{fig:rank}
\end{figure}

As the main goal of these methodologies is to define a set of different symbols $\pi_i \in S_m$ with an unambiguous rule for mapping  between real valued embedded vectors ($X_m$) to these symbols, permutating the ranks according to time, or permutating the time indexes according to the ranks have the same mathematical value.      
Figures \ref{fig:rank} and \ref{fig:order} reveal that the mappings differ in 2 out of 6 symbols, and as $m$ increases so does the differences between both mappings. While this differences between these mappings have no effects in the the calculation of Permutation Entropy with no ties, they play an important role when the alphabet needs to be extended \cite{bian2012modified} to calculate $H(m)$, or when local quantifiers (e.g the Fisher's information) make use of this symbolic dynamics \cite{Martin1999,Vignat2003,PIP228,Rosso2010}. 

\section{Permutation entropy algorithms}\label{PEalg}
The Shannon Entropy of order $m \geq 2$ is defined as 
\begin{equation}\label{eq:PE}
H(m)~=~-\sum_{i=1}^{m!}{p(\pi_i)}{log (p(\pi_i))}.
\end{equation}
When the $p(\pi_i)$ are calculated according the Bandt \& Pompe distribution this Shannon entropy is called the Permutation Entropy \citep{bandt2002permutation}. 
A mapping from $X_m(t)$ to $\pi_i \in ~S_m$ is made for $\forall t \in T$ according to one of the strategies suggested in the previous Section. $\{X_t\}_{t \in T}$ will led to $T-m+1$ symbols ($\pi$) so:
\begin{footnotesize}
\begin{equation}
p(\pi_i)~=~\frac{\sharp \{t|0\leq t \leq T-m,;~ (x_t,x_{t+1},\dots,x_{t+m-1}) \text{, has type }\pi\}}{T-m+1} 
\label{frequ}
\end{equation}
\end{footnotesize}

With the condition $P(x_{t_1}=x_{t_2})=0~\forall t_1 \neq t_2$ all the embedding vectors $X_m(t)$ has $m$ unique values (no ties), and both mappings mentioned in the previous Section lead to the same result of the permutation entropy for each time series $\{X_t\}_{t \in T}$. 

That condition may not comply in several real world time series, so a substantial amount of embedding vectors $X_m(t)$ of these time series could have tied values and the mapping from these $X_m(t)$ to $\pi_i \in ~S_m$ can not be made in neither of the ways presented in Section \ref{OP}. For this reason, different methodologies were developed to handle with time series with tied values.

In essence there are two strategies of dealing with the issue of ties. The first one assumes that the process is indeed continuous so the patterns with ties are in fact missing data deriving from non tied values patterns wrongly observed. The second one makes no such assumption and extends the alphabet in order to ignore the restriction $i_k \neq i_j~ \forall i \neq j$ for $\pi_i=(i_1,i_2,\dots,i_m) \in S_m$ and lets $i_k = i_j $ for $i \neq j$. These types of alphabets have more symbols than $m!$, and varies significantly according if they were constructed with the order based alphabet or the rank based alphabet.
In table \ref{all3patterns} all possible patterns  for $m=3$ are shown in the first column, and their mapping for every methodology are presented.

\subsection{Extended alphabets: Modified Permutation-entropy algorithm}\label{sec:A}
If equal values may represent a feature state of the system under study, mapping equal values in $\{x_t\}$ to equal representation in a symbol $\pi$ could be considered .
\subsubsection{S.1.- Chronological extended alphabet}\label{Bian}
In \cite{bian2012modified} this modified Permutation Entropy is presented.
First like the original order based alphabet, the values of $X_m(t)$ can be sorted in increasing order: $x_{t+i_1-1} \leq x_{t+i_2-1} \leq \dots \leq x_{t+i_m-1}$.
Normally, when there is no equality $x_{t+i_*-1}$ is represented by $i_*$. However, when equality happens, equal values are mapped to the same symbol, which is the smallest index $i_*$ of these equal values, e.g., if  $x_{t+i_{j1}-1} =x_{t+i_{j2}-1}$ and $i_{j1}<i_{j2}$, both $x_{t+i_{j1}-1}$ and $x_{t+i_{j2}-1}$ are represented by $i_{j1}$ in the symbol $\pi_i$. The corresponding permutation symbol of the pattern $X_m(t)$  is defined as: $\pi_i=(i_1,i_2,\dots,i_{j1},i_{j1},\dots,i_m)$.

For example, lets take the series: 
\begin{equation}
X_t=(2,5,1,2,7,1,1,3,1) ~~T=9
\end{equation}

and take the vector $X_5(1)=(2,5,1,2,7)$ this led to the symbol $\pi_i=(31125)$. $X_3(6)=(1,1,3)$ and $X_3(7)=(1,3,1,1)$ map into the symbol $\pi_i=(113)$ and $pi_i=(1112)$ respectively. This modified order based alphabet results in more possible symbols for each embedding dimension $m$ so it characterizes more system states than the original PE method.
\subsubsection{S.2.- Rank extended alphabet}
Again like the original rank based alphabet, $X_m(t)=(x_t,x_{t+1},\dots,x_{t+m-1})$ can be mapped onto $\pi_i=\left(R(x_{t+1}),R(x_{t+2}),\dots,R(x_{t+m}))\right) \in S_m$. If ties exist, those observations get the same rank, that is the lowest rank of all that equal values.
We will take the above time series as an example.
Again take the vector $X_5(1)=(2,5,1,2,7)$ this led to the symbol $\pi_i=(24125)$. $X_3(6)=(1,1,3)$ and $X_3(7)=(1,3,1,1)$ map into the symbol $\pi_i=(113)$ and $\pi_i=(1411)$ respectively. This modified rank based alphabet results in more possible symbols for each embedding dimension $m$ than the original rank alphabet and even more possible symbols than the order based extended alphabet, see Table \ref{table:cantsymbols}.

\begin{table}[]
\centering
\begin{tabular}{|c|c|c|c|c|}
\hline
                              & m=3 & m=4 & m=5 & m=6  \\ \hline
Regular alphabet              & 6   & 24  & 120 & 720  \\ \hline
Chronological extended alphabet & 13  & 73  & 501 & 4051 \\ \hline
Rank extended alphabet  & 13  & 75  & 541 & 4683 \\ \hline
\end{tabular} \caption {\label{table:cantsymbols}Number of symbols $\pi_i$ per embedding dimension $m$ for each alphabet studied in this paper. While for the regular alphabet this quantity is always $m!$, the number of symbols of the order based extended alphabet is much larger(See \citep{bian2012modified} for the explicit formula). The rank based extended alphabet exceed in quantity both alphabets.}
\end{table}

\subsection{Imputation for missing data algorithms}\label{sec:B}
If one assumes that the process is indeed of continuous data then the patterns with ties are in fact missing data deriving from non tied values patterns wrongly observed.
Missing data are a common problem in all types of research fields and there are various methods of handling with missing data, that can be divided into two major strategies: the first is eliminate all the observation that are incomplete (and work only with the complete cases), and the second one is to impute the missing data with a suitable value\cite{Donders2006}. For all the following examples the rank based alphabet will be used.
\subsubsection{S.3.- Complete Cases}\label{CC}
This method was originally suggested in \cite{bandt2002permutation} and it is analogue to a complete case analysis in the missing value statistic theory . It is simply to eliminate the patterns that contain tied values. 
For example for $m=3$ the series $X_t=(2,5,1,2,7,1,1,3,1)$ and $m=3$, only the patterns $X_3(1)=(2,5,1)$, $X_3(2)=(5,1,2)$, $X_3(3)=(1,2,7)$, $X_3(4)=(2,7,1)$ are mapped onto $\pi=(231),\pi=(312),\pi=(123),\pi=(231)$ respectively and  $X_3(5)=(7,1,1)$, $X_3(6)=(1,1,3)$, $X_3(7)=(1,3,1)$ are eliminated.
Then compute the $p^*(\pi_i)$ for all the remaining vectors.
\subsubsection{S.4.- Time Ordered Imputation}
This is one of the most used techniques used to deal with repeated values. It is used in \cite{MatillaGarcía2014,parlitz2012classifying,Cao2004,Saco2010,Zunino2008} among others. It simply states that if $x_{t_1}=x_{t_2}$ and $t_1<t_2$ then $x_{t_1}<x_{t_2}$.
Following the previous example, $X_3(1)=(2,5,1)$, $X_3(2)=(5,1,2)$, $X_3(3)=(1,2,7)$, $X_3(4)=(2,7,1)$ are equally mapped since they do not have equal values and  $X_3(5)=(7,1,1)$ is mapped to $\pi=(312)$, $X_3(6)=(1,1,3)$ to $\pi=(123)$ ,and finally  $X_3(7)=(1,3,1)$ is mapped to $\pi_i=(132)$. 
Then compute the resulting $p^*(\pi_i)$.

\subsubsection{S.5.- Random Imputation}
In \citep{bandt2002permutation}, they recommend to numerically break equalities by adding small perturbations. The amplitude of this perturbation should be less than the minimum difference between two different values of the pattern.
This means, by our example, that $X_3(5)=(7,1,1)$ should be mapped either to $\pi=(312)$ or $\pi=(321)$, with probability $1/2$ each, since the perturbation is white noise, but never to other symbol than those because no matter the result of the random perturbation the first value should have always the highest rank. Following the example, $X_3(6)=(1,1,3)$ is going to mapped either to $\pi=(123)$ or $\pi=(213)$ and finally  $X_3(7)=(1,3,1)$ is mapped either to $\pi_i=(132)$ or $\pi=(231)$ with the same probability.
For instance, for $m=3$, if all three values of the pattern are equal (e.g $X_3(t_k)=(7,7,7)$, any one of the six symbols could appear, each with probability $1/6$.
Then compute the resulting $p^*(\pi_i)$.

\subsubsection{S.6.- Bayesian Imputation}\label{BayImp}
Random Imputation suggests that independently of the  time series in study, as patterns with equal values are the result of a coarse observation of patterns without ties in the original series, they should be mapped to the symbols correspondent to those original patterns with the same probability. 
It is shown that in most situations, simple techniques for handling missing data (such as complete cases or random imputation) in other research areas produces biased results \cite{Donders2006}, and there are more sophisticated techniques that give much better results. With these techniques, missing data for a subject are imputed by a value that is predicted using the subject's other, known characteristic.
We propose a method which is similar to the random imputation but instead of adding a random perturbation that maps with equal probabilities to each suitable symbol, this probabilities are originated with a previous known Probability Distribuition Function, and are not necessary equal. The PDF proposed as a prior distribution is the one resulting of computing the $p(\pi_i) \forall i$ for the complete cases. See subsection  \ref{CC}.

There are seven steps to follow:

\begin{enumerate}
\item Define the embedding dimension $m$. That leads to $S_m=\{\pi_i\}$ all the $m!$ possible permutations of $(1,2,\dots,m)$.
\item Map $X_m(t)~ \forall t$ to their correspondent $\pi_i \in Sm$ according to their rank (section \ref{RankBA}).
\item If there is any ties in  $X_m(t)$ for any $t$, eliminate the vector (section \ref{CC}).
\item Calculate the $p^*(\pi_i)$ (equation \ref{frequ}).
\item Repeat the procedure of mapping every $X_m(t)~ \forall t$ to their correspondent $\pi_i \in Sm$ according to their rank, but do not eliminate the vectors $X_m(t)$ with repeated values.
\item For each vector $X_m(t)$ with repeated values do the mapping to a suitable $\pi_i$ but with probability $p^*(\pi_i)$ for each  $\pi_i$.
\item Calculate the new $p(\pi_i)$
\end{enumerate}

\begin{figure}[h!]
\centering
\includegraphics[width=80mm]{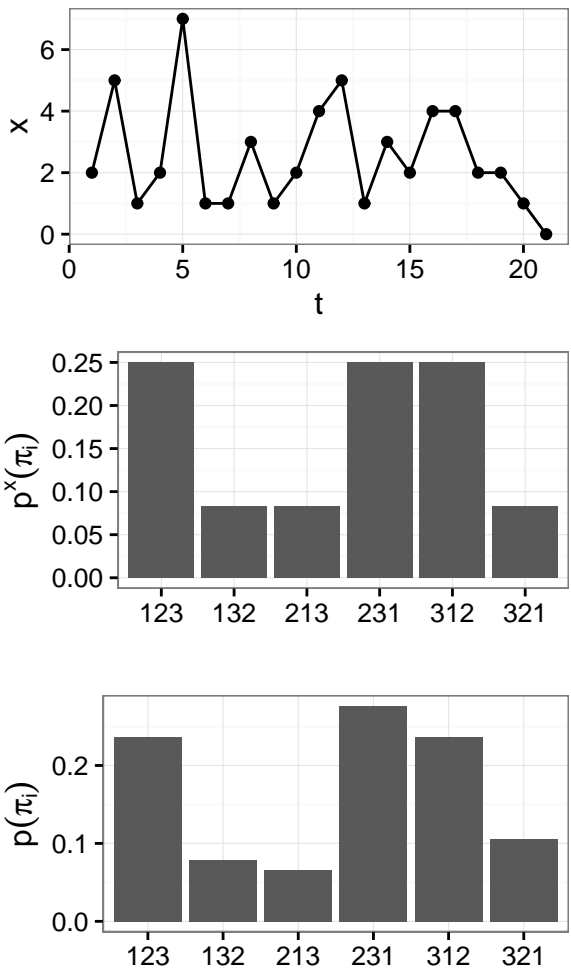}
\caption{The above plot shows the complete time series $X_t=(2,5,1,2,7,1,1,3,1,2,4,5,1,3,2,4,4,2,2,1,0)$, the middle plot computed the Bandt Pompe's PDF eliminating patterns $X_3(t)$ with repeated values for $m=3$, using the remaining 12 complete cases. The third plot shows the resulting PDF with all 19 cases, imputing to the patterns with ties a suitable symbol. But the probability of choosing  that symbol is according to the previously calculated PDF using complete cases methodology.}
\label{Bayesianexample}
\end{figure}

$X_t=(2,5,1,2,7,1,1,3,1,2,4,5,1,3,2,4,4,2,2,1,0)$ is illustrated in Fig. \ref{Bayesianexample} as an example of the Bayesian Imputation. Table \ref{Comp} maps every pattern $X_3(t)$ of this time series for all the methodologies listed on this contribution.

\begin{table}[]
\centering
\label{Comp}
\begin{tabular}{ccc|c|c|c|c|c|c|}
\cline{4-9}
                        &                        &   & \multicolumn{6}{c|}{$\pi_i$}                                                                                                                                                                                                                                                                                                                        \\ \hline
\multicolumn{3}{|c|}{$X_3(t)$}                       & \begin{tabular}[c]{@{}c@{}}Chron.  \\ Ext.\end{tabular} & \begin{tabular}[c]{@{}c@{}}Rank \\ Ext.\end{tabular} & \begin{tabular}[c]{@{}c@{}}Comp.\\ Cases\end{tabular} & \begin{tabular}[c]{@{}c@{}}Time\\ Ord.\end{tabular} & \begin{tabular}[c]{@{}c@{}}Rand.\\ Imp.\end{tabular} & \begin{tabular}[c]{@{}c@{}}Bayesian \\ Imp.\end{tabular} \\ \hline
\multicolumn{1}{|c|}{2} & \multicolumn{1}{c|}{5} & 1 & 312                                                          & 231                                                        & 231                                                & 231                                                 & 231                                                  & 231                                                   \\ \hline
\multicolumn{1}{|c|}{5} & \multicolumn{1}{c|}{1} & 2 & 231                                                          & 312                                                        & 312                                                & 312                                                 & 312                                                  & 312                                                   \\ \hline
\multicolumn{1}{|c|}{1} & \multicolumn{1}{c|}{2} & 7 & 123                                                          & 123                                                        & 123                                                & 123                                                 & 123                                                  & 123                                                   \\ \hline
\multicolumn{1}{|c|}{2} & \multicolumn{1}{c|}{7} & 1 & 312                                                          & 231                                                        & 231                                                & 231                                                 & 231                                                  & 231                                                   \\ \hline
\multicolumn{1}{|c|}{7} & \multicolumn{1}{c|}{1} & 1 & 221                                                          & 311                                                        & X                                                  & 312                                                 & \begin{tabular}[c]{@{}c@{}} 312 $\tiny{p=\frac{1}{2}}$\\ 321 $ \tiny{p=\frac{1}{2}}$\end{tabular}   & \begin{tabular}[c]{@{}c@{}} 312 $\tiny{p^*=\frac{1}{4}}$\\ 321 $\tiny{p^*=\frac{1}{12}}$\end{tabular}     \\ \hline
\multicolumn{1}{|c|}{1} & \multicolumn{1}{c|}{1} & 3 & 113                                                          & 113                                                        & X                                                  & 123                                                 & \begin{tabular}[c]{@{}c@{}} 123 $ \tiny{p=\frac{1}{2}}$\\ 213 $ \tiny{p=\frac{1}{2}}$\end{tabular}    & \begin{tabular}[c]{@{}c@{}} 123 $\tiny{p^*=\frac{1}{4}}$\\ 213 $\tiny{p^*=\frac{1}{12}}$\end{tabular}     \\ \hline
\multicolumn{1}{|c|}{1} & \multicolumn{1}{c|}{3} & 1 & 112                                                          & 131                                                        & X                                                  & 132                                                 & \begin{tabular}[c]{@{}c@{}} 132 $ \tiny{p=\frac{1}{2}}$\\ 231 $ \tiny{p=\frac{1}{2}}$\end{tabular}    & \begin{tabular}[c]{@{}c@{}} 132 $\tiny{p^*=\frac{1}{12}}$\\ 231 $\tiny{p^*=\frac{1}{4}}$\end{tabular}     \\ \hline
\multicolumn{1}{|c|}{3} & \multicolumn{1}{c|}{1} & 2 & 231                                                          & 312                                                        & 312                                                & 312                                                 & 312                                                  & 312                                                   \\ \hline
\multicolumn{1}{|c|}{1} & \multicolumn{1}{c|}{2} & 4 & 123                                                          & 123                                                        & 123                                                & 123                                                 & 123                                                  & 213                                                   \\ \hline
\multicolumn{1}{|c|}{2} & \multicolumn{1}{c|}{4} & 5 & 123                                                          & 123                                                        & 123                                                & 123                                                 & 123                                                  & 123                                                   \\ \hline
\multicolumn{1}{|c|}{4} & \multicolumn{1}{c|}{5} & 1 & 312                                                          & 231                                                        & 231                                                & 231                                                 & 231                                                  & 231                                                   \\ \hline
\multicolumn{1}{|c|}{5} & \multicolumn{1}{c|}{1} & 3 & 231                                                          & 312                                                        & 312                                                & 312                                                 & 312                                                  & 312                                                   \\ \hline
\multicolumn{1}{|c|}{1} & \multicolumn{1}{c|}{3} & 2 & 132                                                          & 132                                                        & 132                                                & 132                                                 & 132                                                  & 132                                                   \\ \hline
\multicolumn{1}{|c|}{3} & \multicolumn{1}{c|}{2} & 4 & 213                                                          & 213                                                        & 213                                                & 213                                                 & 213                                                  & 213                                                   \\ \hline
\multicolumn{1}{|c|}{2} & \multicolumn{1}{c|}{4} & 4 & 122                                                          & 122                                                        & X                                                  & 123                                                 & \begin{tabular}[c]{@{}c@{}} 123 $ \tiny{p=\frac{1}{2}}$\\ 132 $ \tiny{p=\frac{1}{2}}$\end{tabular}    & \begin{tabular}[c]{@{}c@{}} 123 $\tiny{p^*=\frac{1}{4}}$\\ 132 $\tiny{p^*=\frac{1}{12}}$\end{tabular}     \\ \hline
\multicolumn{1}{|c|}{4} & \multicolumn{1}{c|}{4} & 2 & 311                                                          & 221                                                        & X                                                  & 231                                                 & \begin{tabular}[c]{@{}c@{}} 231 $\tiny{p=\frac{1}{2}}$\\ 321 $\tiny{p=\frac{1}{2}}$\end{tabular}    & \begin{tabular}[c]{@{}c@{}} 231 $\tiny{p^*=\frac{1}{4}}$\\ 321 $\tiny{p^*=\frac{1}{12}}$\end{tabular}     \\ \hline
\multicolumn{1}{|c|}{4} & \multicolumn{1}{c|}{2} & 2 & 221                                                          & 311                                                        & X                                                  & 312                                                 & \begin{tabular}[c]{@{}c@{}} 312 $ \tiny{p=\frac{1}{2}}$\\ 321 $\tiny{p=\frac{1}{2}}$\end{tabular}    & \begin{tabular}[c]{@{}c@{}} 312$\tiny{p^*=\frac{1}{4}}$\\ 321 $\tiny{p^*=\frac{1}{12}}$\end{tabular}     \\ \hline
\multicolumn{1}{|c|}{2} & \multicolumn{1}{c|}{2} & 1 & 311                                                          & 221                                                        & X                                                  & 231                                                 & \begin{tabular}[c]{@{}c@{}}231 $ \tiny{p=\frac{1}{2}}$\\ 321 $ \tiny{p=\frac{1}{2}}$\end{tabular}    & \begin{tabular}[c]{@{}c@{}} 231 $\tiny{p=^*\frac{1}{4}}$\\ 321 $\tiny{p^*=\frac{1}{12}}$\end{tabular}     \\ \hline
\multicolumn{1}{|c|}{2} & \multicolumn{1}{c|}{1} & 0 & 321                                                          & 321                                                        & 321                                                & 321                                                 & 321                                                  & 321                                                   \\ \hline
\end{tabular}\caption{ Different mappings for the example  $X_t=(2,5,1,2,7,1,1,3,1,2,4,5,1,3,2,4,4,2,2,1,0)$. Extending the alphabet implies that repeated values in $X_3(t)$ are patterns that represent the nature of the process, so they are represented with a symbol $\pi_i$ with some equal elements. The four methodologies on the right assume that $X_3(t)$ with repeated values are a particular case of missing data. \textit{Complete Cases} removes that patterns and calculate $p*(\pi_i)$ without them. Time Ordered Imputation assumes that the value that first appear is the lower valued. At last, Random Imputation and Bayesian Imputation replace the missing data element with a suitable value, the first imputes a random suitable symbol while the second takes account of the original structure using the probabilities $p^*(\pi_i)$ of Complete Cases for imputing the value.}
\end{table}

\section{Numerical Results}\label{results} 
In this Section the above strategies for missing data imputation are evaluated using data from simulated chaotic processes. In order to get a reproducible set of time series, all maps presented in \cite{Rosso2013} were simulated using the initial conditions presented therein. All those series presented none, or negligible amount, patterns $X_m(t)$ with ties. These time series will be referred as the original time series. 
After that, each original time series was truncated up to one decimal resolution, leading to a coarse version of those original time series. Due to that finite resolution, this coarse versions have an amount of patterns with tied values that can not be considered negligible.
The simulation consists in 39 time series generated by different chaotic processes, and each one was simulated for a length of $n=(5000,10000,30000,90000)$, all those series presented none, or a negligible amount, patterns $X_m(t)$ with ties, and for each series the Permutation Entropy, $H(m)$, was calculated for embedding dimension $m=\{3,4,5,6\}$. For those initial condition, as the series is deterministic, the $H(m)~,~m=\{3,4,5,6\}$ are the actual Permutation Entropies of the process.
Next, these series are truncated to one decimal, so the amount of patterns $X_m(t)$ with ties are not negligible any more. All the strategies enunciated in Sections \ref{sec:A} and \ref{sec:B} are used separately for each truncated series to compute the $p(\pi_i)$, and their respective $\hat{H}(m)~,~ m=\{3,4,5,6\}$.

This yields to quantify how well each strategy estimates the actual Permutation Entropy ($H(m)$) comparing the result of each original time series with their respective Permutation Entropy calculated with all the above methodologies over the coarse version.\\

\begin{figure*}[h!]
\includegraphics[width=190mm]{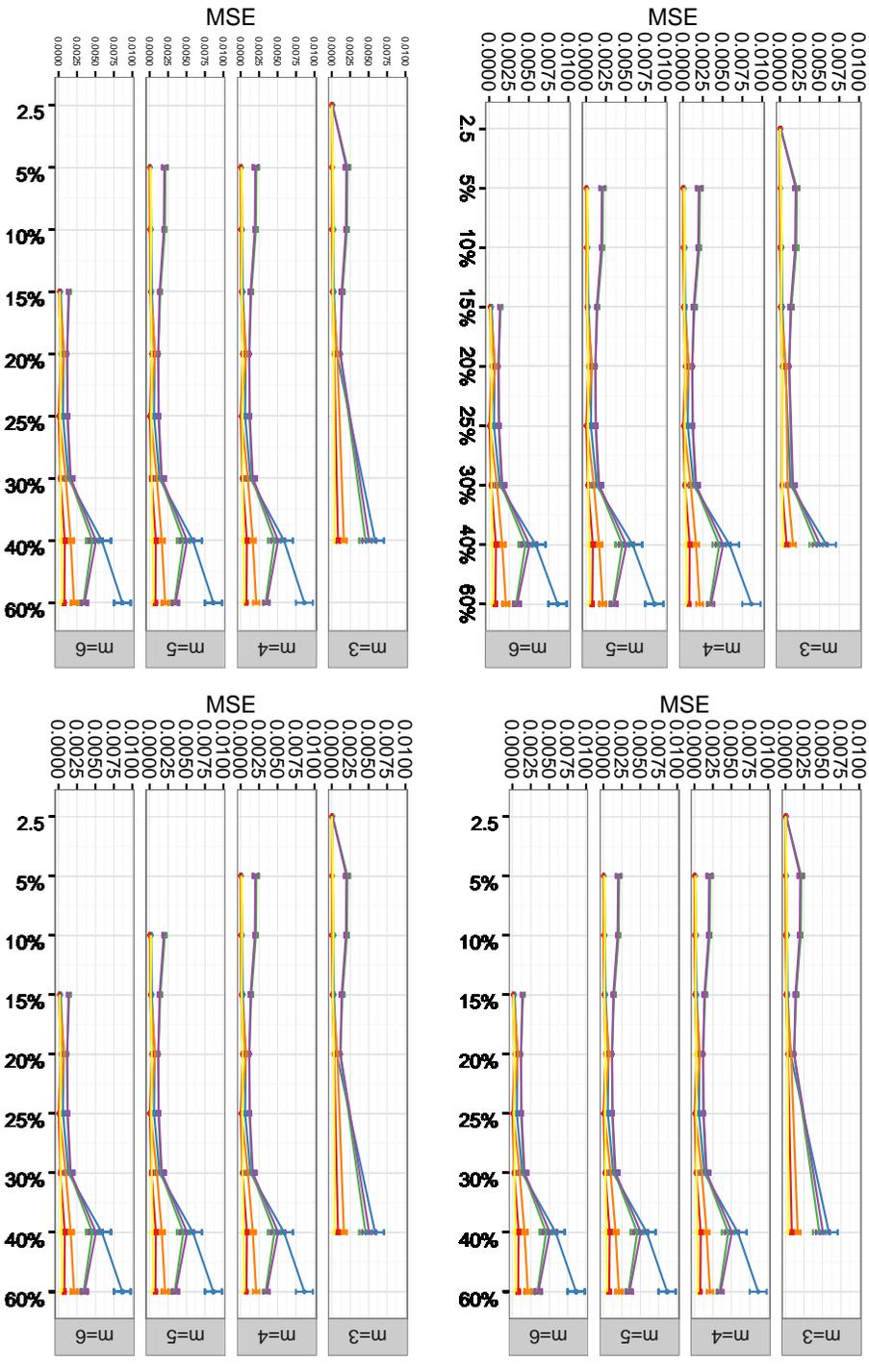}
\caption{\label{fig:MSE}The MSE is plotted for different ratios of patterns with ties for every methodology enunciated in section \ref{PEalg} and for every length of the simulated time series \textit{($n=5000,10000,30000,90000$ begining in the top and to the right)} . The color code is as follows: red for Bayesian Imputation, blue for Random Imputation, green for the Chronological Extended Alphabet, purple por the Rank Extended Alphabet, orange for the Complete Cases methodology and finally yellow for the Time Ordered Imputation. Bayesian Imputation and Time Order Imputation outperform all the other methodologies independently of the embedding dimension $m$ along every rate of patterns with ties. This figure shows that the estimation for each methodology does not depend on the length of the series}
\end{figure*}

\begin{figure}[h!]
\includegraphics[width=90mm]{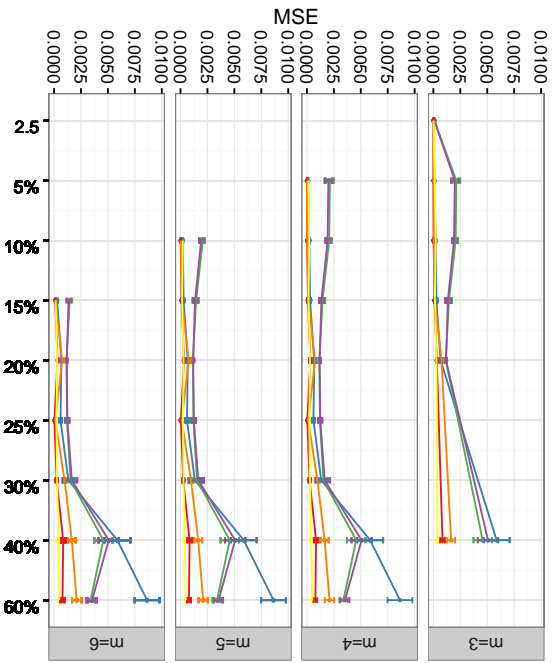}
\caption{\label{fig:n90000}\textbf{$n=90000$} The MSE is plotted for different ratios of patterns with ties (i.e. \# of patterns with tied values over total amount of patterns) for every methodology enunciated in section \ref{PEalg}. The color code is as follows: red for Bayesian Imputation, blue for Random Imputation, green for the Chronological Extended Alphabet, purple por the Rank Extended Alphabet, orange for the Complete Cases methodology and finally yellow for the Time Ordered Imputation. Bayesian Imputation and Time Order Imputation outperform all the other methodologies independently of the embedding dimension $m$ along every ratio of patterns with ties.}
\end{figure}

In Figure \ref{fig:MSE} the results for each length $n$ of the series is shown faceting with the embedding dimension $m$. The truncated series were grouped by their percentage of patterns $\{X_m(t)\}$ with ties over the all the patterns (i.e. the ratio of missing values over the total, this ratio shall be referred to as 'repeated ratio' this point onwards). To quantify the behaviour of the estimator, the Mean Square Error, defined as $E\left[(\hat{H}(m)-H(m))^2\right]$, is used.In each figure the Mean Square Error (MSE) was plotted against those groups for every embedding dimension $m$. The repeated ratio depends not only of the structure of the time series but also of the embedding dimension $m$. Figure \ref{fig:n90000} focus on the particular case of $n=90000$. 

\begin{figure*}[h!]
\includegraphics[width=190mm]{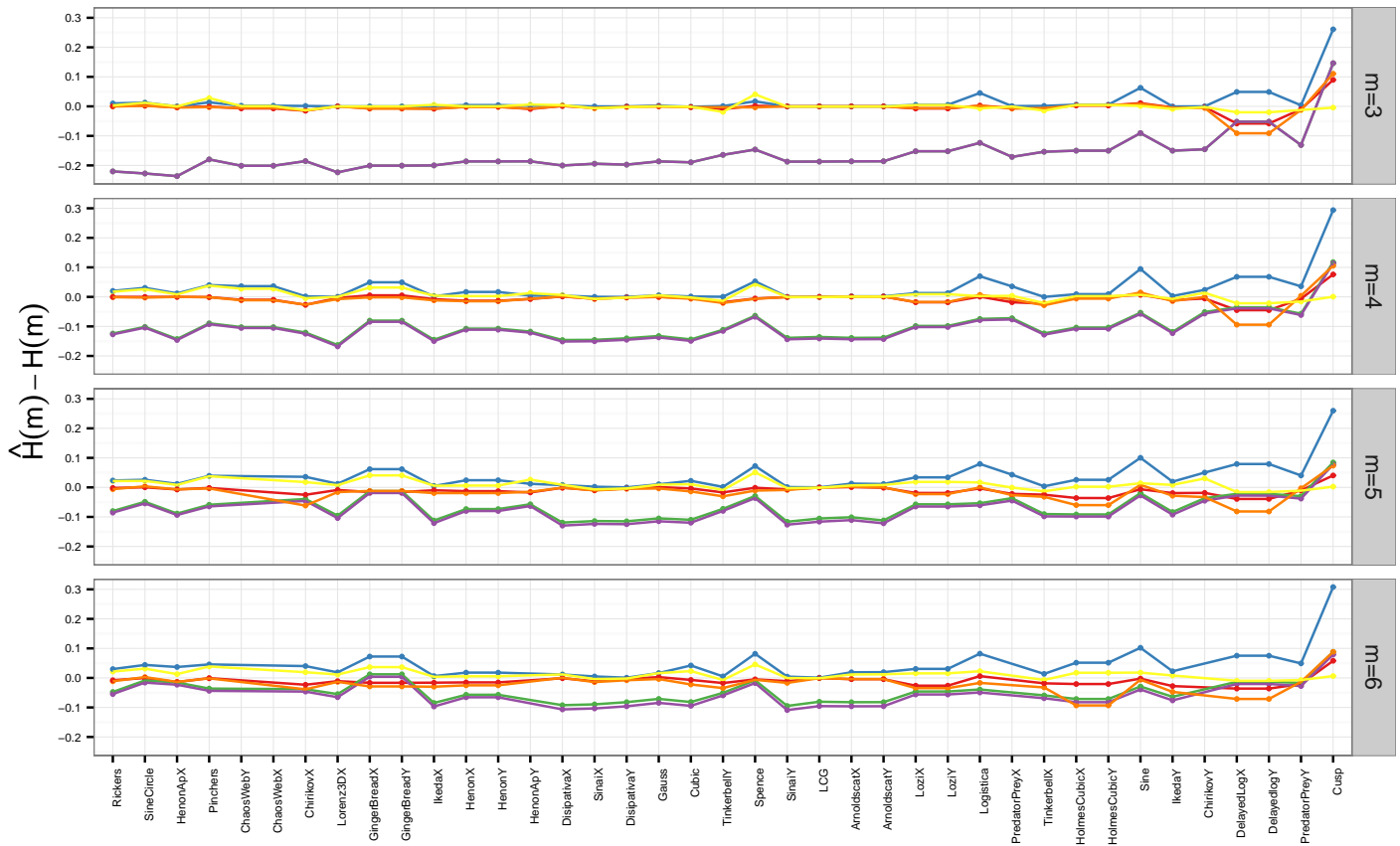}
\caption{\label{fig:Error} The difference between the estimation and the real value $\hat{H}(m) - H(m)$, for $n=90000$ is shown. The color code is as follows: red for Bayesian Imputation, blue for Random Imputation, green for the Chronological Extended Alphabet, purple por the Rank Extended Alphabet, orange for the Complete Cases methodology and finally yellow for the Time Ordered Imputation. The different maps in are ordered in increasing amount of repeated values. Extending alphabet methodologies tend to sub estimate the entropy, specially when the embedding dimension is small. Random Imputation methodology over estimate the entropy as it create noise, this is more evident when the imputation is made over a  large amount of values. Time Ordered Imputation and Bayesian Imputation have similar performance, and both outperforms Complete Cases Methodology }
\end{figure*}

\begin{figure*}[h!]
\includegraphics[width=180mm]{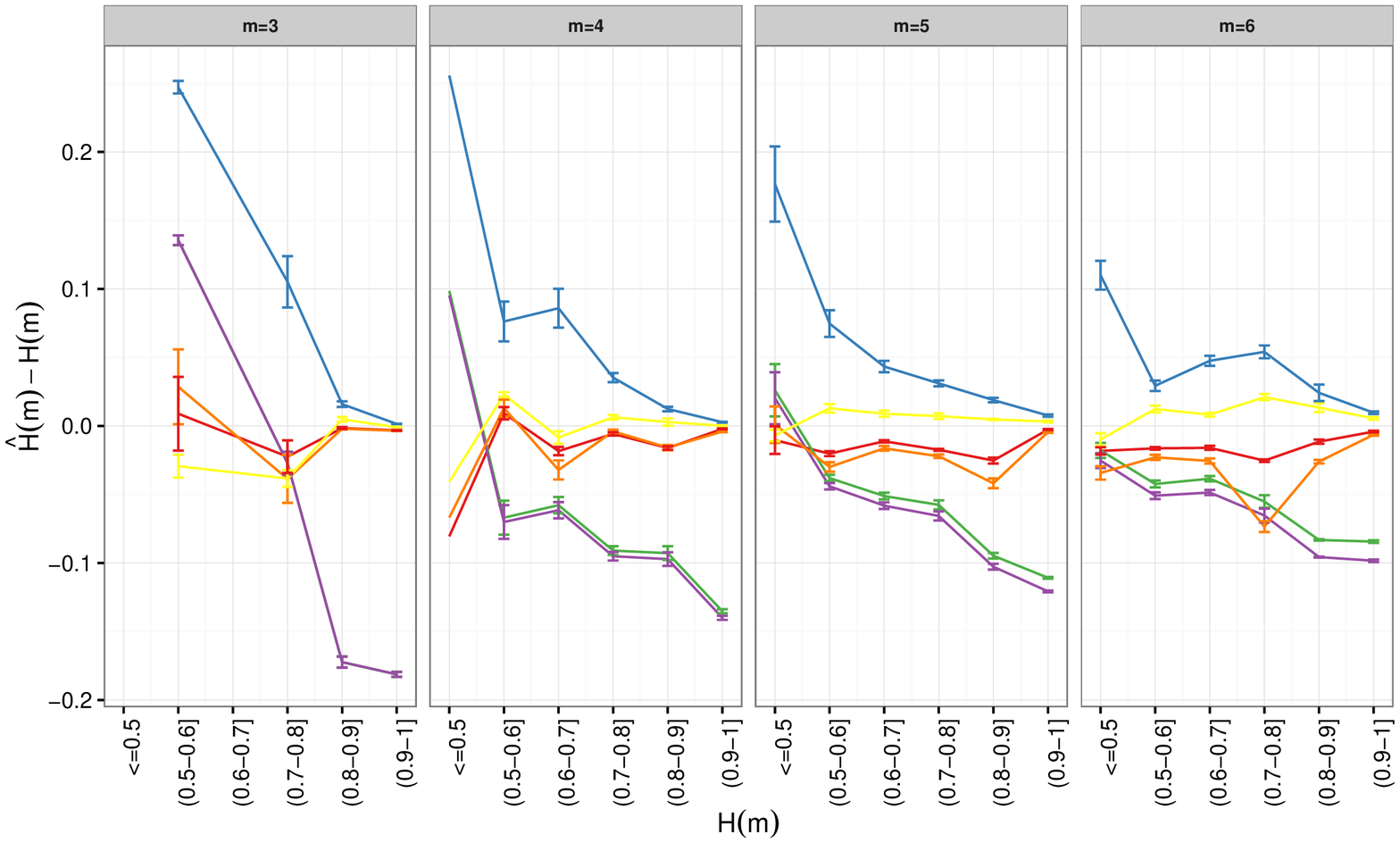}
\caption{\label{fig:errornivel} The difference between the estimation and the real value is plotted. Time series were grouped by the level of their Permutation Entropy in the original process, and the error $\hat{H}(m)-H(m)$, for every methodology  was calculated. As stated before, Random Imputation proposed in the seminal paper always increases the value of the estimator $\hat{H}(m)$, and overestimate $H(m)$. It is much more noticeable when the $H(m)$ of the original series is low valued.
Extending the alphabet always underestimate the Permutation Entropy, doing a poorly job for High values of $H(m)$. In Complete Cases, $\hat{H}(m)$ underestimates low values of $H(m)$, but when the $H(m)$ is high is not biased. This is even more evident for the case $m=6$. Bayesian Imputation and Time Ordered Imputation outperforms the other methodologies for every level of $H(m)$.}
\end{figure*}

Time Order Imputation and Bayesian Imputation consistently beat the other methodologies for every embedding dimension $m$ and for all the repeated ratios as the MSE is the lowest in all the cases.  
As for the Bias of the estimation for each algorithm Fig. \ref{fig:Error} shows the difference between the estimation and the real value, $\hat{H}(m)-H(m)$, for $n=90000$. The different chaotic maps are ordered by increasing amount of repeated values. Methodologies that use extended alphabet tend to sub estimate the entropy, specially when the embedding dimension is small. Random Imputation methodology over estimates the entropy as it create noise and this is more evident when the imputation is made over a  time series with a large number of repeated values. Time Ordered Imputation and Bayesian Imputation have similar performance, and both outperform Complete Cases Methodology.
In Fig.\ref{fig:errornivel}, time series were grouped by the level of their Permutation Entropy in the original process, and the error $\hat{H}(m)-H(m)$, for every methodology  was calculated. As stated before, Random Imputation proposed in the seminal paper always increases the value of the estimator $\hat{H}(m)$, and overestimate $H(m)$. It is much more noticeable when the $H(m)$ of the original series is low valued.
Extending the alphabet always underestimate the Permutation Entropy, doing a poorly job for High values of $H(m)$. In Complete Cases, $\hat{H}(m)$ underestimates low values of $H(m)$, but when the $H(m)$ is high is not biased. This is even more evident for the case $m=6$. Bayesian Imputation and Time Ordered Imputation outperforms the other methodologies for every level of $H(m)$.

\section{Real Application}\label{realAP}
We applied the presented strategies to classify using heart rate  time series from a (healthy) control group and from patients suffering from Congestive Heart Failure (CHF). The data have been collected from internet data bases: NSR2DB \textit{Normal Sinus Rhythm RR interval database} for the healthy patients and CHF2DB \textit{Congestive heart failure RR interval database}, for the patients suffering from CHF, from http://\texttt{www.physionet.org/cgi-bin/atm/ATM}.
For each database, 15 series were taken, each one with approximately 100000 observations. In HRV series (RRintervals i.e the interval from the peak of one QRS complex to the peak of the next one as shown on an electrocardiogram (ECG)) the case of equal values appears very frequently due to limited sampling frequency of the ECG. So due to the high frequency of equal values in HRV series, these are good examples for testing the methodologies presented in Sections \ref{sec:A} and \ref{sec:B}.

\begin{figure*}[h!]
\includegraphics[width=180mm]{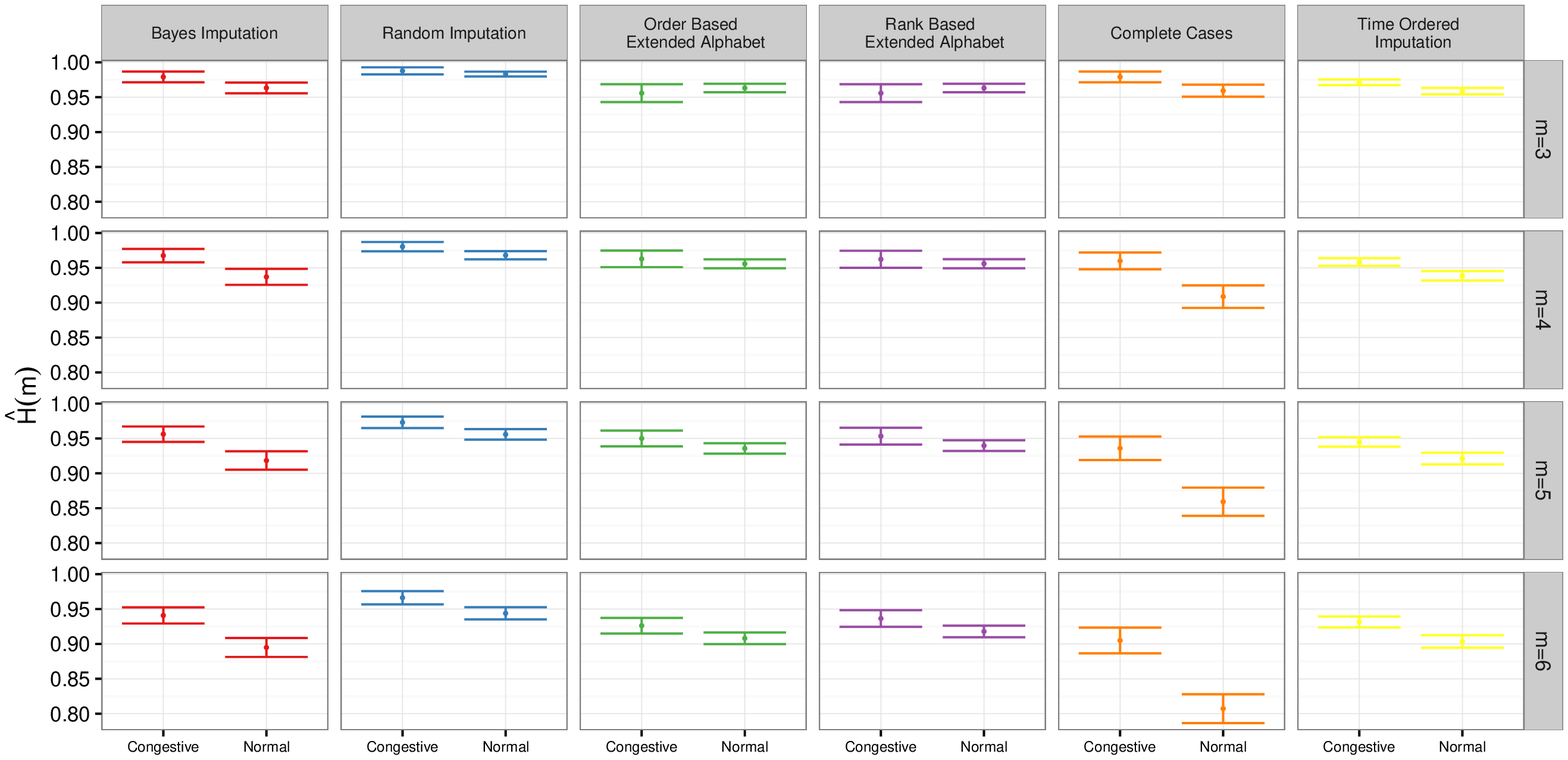}
\caption{\label{fig:ECG} The mean of the calculated Permutation Entropies $\hat{H}(m)$ are plotted for each type of patient, those suffering from CHF (i.e. Congestive) and those who do not (i.e. Normal) along with its standard error. The plot is divided by each methodology and for every embedding dimension $m =\{3,4,5,6\}$. In this sample Complete Cases works fine separating the means, but it can result in  several misclassification because of the shape of the PDF of each group (See \ref{fig:ECGbox}). Random Imputation also separates the means of both groups, but suffers from the same problematic. Extending the alphabet does not perform well as a classifier in this problematic. Bayesian Imputation and Time Ordered Imputation are excellent classifiers  since  they do great job separating the mean populations.}
\end{figure*}

\begin{figure*}[h!]
\includegraphics[width=180mm]{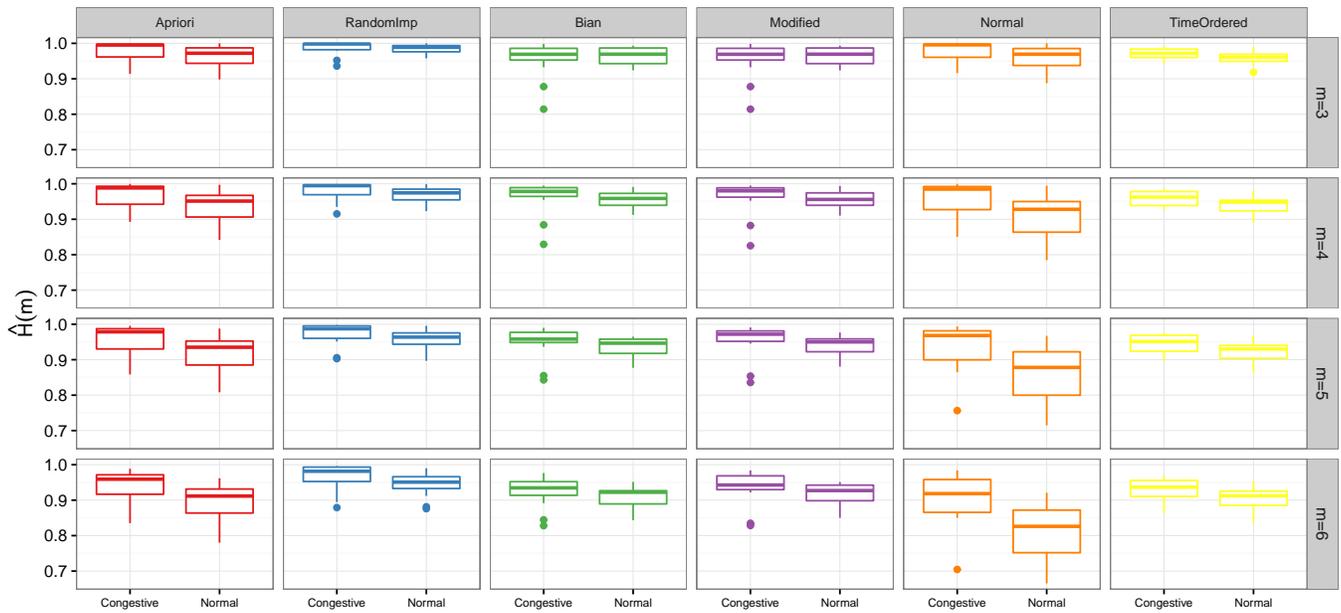}
\caption{\label{fig:ECGbox} A boxplot for each type of patient, those suffering from CHF (i.e. Congestive) and those who do not (i.e. Normal) is drawn, divided by each methodology and for every embedding dimension $m =\{3,4,5,6\}$ so not only the mean and the dispersion is noticed but also the shape. In this sample Complete Cases works fine, but it can result in  several misclassification, labelling as normal patients the ones suffering from Congestive Heart Failure due to the artificiality long tailed distribution of this last group. Random Imputation also suffers form the different shapes on the probability distribution function of the groups so does not perform well as a classifier . Extending the alphabet does not distinguish between the two populations. Bayesian Imputation is the optimum classifier  since it does a great job separating the populations and obtains the lowest p-values in the test (table \ref{table:pvalues}. }
\end{figure*}

In fig. \ref{fig:ECG} the mean of the calculated Permutation Entropies $\hat{H}(m)$ are plotted for each type of patient, those suffering from CHF (i.e. Congestive) and those who do not (i.e. Normal) along with its standard error. The plot is divided by each methodology and for every embedding dimension $m =\{3,4,5,6\}$.
In fig. \ref{fig:ECGbox} the boxplot of the calculated Permutation Entropies are plotted for each type of patient, so the shape and outliers, among other features can be appreciated.
 A Mann-Whitney test was conduced in all those cases for assessing the difference between the two populations. This test can detect differences in shape and spread as well as just differences in medians. The resultant p-values are shown in table \ref{table:pvalues}. 
Eliminating the patterns with repeated values (Complete Cases) is the methodology that best separates the mean but will not work well as a classifier. As can be seen in Fig. \ref{fig:errornivel}, this methodology underestimates the Permutation Entropy for low levels of $H(m)$ but has no bias when the $H(m)$ is near one. 
In this sample Complete Cases works fine, but it can result in several misclassification, labelling as normal patients the ones suffering from Congestive Heart Failure due to the artificially long tailed distribution, as can be seen in \ref{fig:ECGbox}, generated by this relation within the $H(m)$ level and it bias.
Random Imputation also separates the mean but suffers form this relation too. However in this case, $\hat{H}(m)$ overestimate $H(m)$ when it is low but this does not happen for high levels of $H(m)$, so it would result in a misclassification, labelling as \textit{Congestive} the patients that are healthy.
Extending the alphabet does not perform well in this problematic. 
Bayesian Imputation is the optimum classifier for this application since it does a great job separating the populations (see table \ref{table:pvalues})  and the bias does not depend of the level of the estimated Permutation Entropy, so if a misclassification occurs it would not be skewed to either population.

\begin{table*}

    \begin{tabular}{|c|c|c|c|c|c|c|}
        \hline
     \begin{tabular}[c]{@{}c@{}} Embedding\\ Dimension\end{tabular} & \begin{tabular}[c]{@{}c@{}}Bayesian Imputation\end{tabular} & \begin{tabular}[c]{@{}c@{}}Random Imputation\end{tabular} & \begin{tabular}[c]{@{}c@{}} Chronological \\ Extended Alphabet\end{tabular} &  \begin{tabular}[c]{@{}c@{}} Rank  \\ Extended Alphabet\end{tabular} &                                                            \begin{tabular}[c]{@{}c@{}} Complete Cases\end{tabular} & \begin{tabular}[c]{@{}c@{}} Time Ordered \\ Imputation\end{tabular} \\ \hline
        3                   & 0.0502           & 0.0555            & 0.7437                        & 0.7437                       & 0.0408         & 0.0555                  \\ \hline
        4                   & 0.0295           & 0.0502            & 0.0742                        & 0.0814                       & 0.0185         & 0.0975                  \\ \hline
        5                   & 0.0209           & 0.0329            & 0.0742                        & 0.0555                       & 0.0036         & 0.1261                  \\ \hline
        6                   & 0.0086           & 0.0234            & 0.0814                        & 0.0555                       & 0.0009         & 0.0674                  \\
        \hline
    \end{tabular}
\caption{\label{table:pvalues} The p-values of Fig. \ref{fig:ECGbox} are presented.\\ A Mann-Whitney test was conduced in all those cases for assessing the difference between the two populations. This test can detect differences in shape and spread as well as just differences in medians. Bayesian Imputation is the optimum classifier  since it does a great job separating the populations and obtains the lowest p-values.}\end{table*}

\section{Conclusions}\label{Conclusions}

When quantifying complexity for a given time series $\{X_t\}_{t \in T}$ entropy measures are an excellent choice, but common techniques neglect any effects stemming from the temporal order of the values $x_i~ \text{in}~ \{X_t\}$. In order to take account this casual information, the time series can be encoded into a  sequences of symbols. Bandt \& Pompe, in  \cite{bandt2002permutation} proposed a natural encoding by partitioning (a non disjoint partition) $\{X_t\}_{t \in T}$ in vectors $X_t(m)$ of lenght $m$, called the embedding dimension, for all $t \in T$ and mapping those $m$-dimensional vectors (i.e patterns) onto symbols ($\pi_i=(i_1,i_2,\dots,i_m)~~i_j \neq i_k~\forall j\neq k$), based on the theory of symbolic dynamics.
As mentioned in section \ref{OP}, two different mapping to the symbolic alphabet are used in the literature concerning the Permutation Entropy for constructing the $\pi_i$ and the subsequent $p(\pi_i)$ that are used to calculate $H(m)$, by permutating the time index (i.e\textit{Chronological Alphabet Mapping}) and by permutating the ranks (i.e \textit{Rank Alphabet Mapping}). The first one, \textit{the Chronological}, rearrange the time indexes according to their amplitude, and the second one \textit{the Rank} rearrange the amplitude indexes (ranks) according to their location in time. While for computing $H(m)$ there is no difference in choosing either one of the alphabets (as long the restriction $i_j \neq i_k~\forall j\neq k$ holds $\forall \pi_i$), when the desire is to apply the approach of Bandt \& Pompe to local quantifiers the election of the alphabet matters and further research must be made.\\
Another issue is that this original approach assumed that $x_t$ in $\{X_t\}_{t \in T}$ has a continuous probability distribution function $\forall t \in T$, thus equal values appear with probability 0. In a variety of time series stemming of real life processes, e.g. HRV series, equal values in a pattern $X_t(m)$ appear often and can not be neglected without consequences. This increase in the frequency of patterns with equal values may occur by two major reasons: either the realization is not from a process with a continuous probability distribution function, so the repeated values represent the dynamics of the process, or indeed the assumptions of the process are complied but repeated values appear with high frequency due to a limited resolution in the data collection process.\\
In order to deal with this issue, \citet{bian2012modified} proposed an extended alphabet that would take account of the repeated values on the symbolic representation $\pi_i=(i_1,i_2,\dots,i_m)$ by ignoring the restriction $i_j \neq i_k~\forall j\neq k$ (see section \ref{Bian}). Various points should be considered in relation to this methodology. \\
First of all if the repeated values of the time series  $\{X_t\}$ under study are supposed to be due to low resolution, the augmented number of states (see Table \ref{table:cantsymbols}) does not represent the states of the process. Even more, if the repeated ratio is not large enough, this methodology will greatly underestimate the real Permutation Entropy of the process because there will much more states with little representation, not because of the nature of the process but because of the fictitious states incorporated by the extension of the alphabet.\\
As shown recently by Amig\'o et al.\cite{Amigo2007, Amigo2008}, in the case of deterministic one-dimensional maps not all the possible
ordinal patterns can be effectively materialized into orbits,
which in a sense makes these patterns ''forbidden''. Indeed, the existence of these forbidden ordinal patterns becomes a
persistent dynamical property. That is, for a fixed pattern length (embedding dimension) $m$ the number of forbidden patterns
present in the time series (unobserved patterns) is independent of the series length T \cite{Carpi2010} and has a strong relation with the Permutation Entropy.
In fact, the methodology used in \citet{bian2012modified} increases the number of forbidden patterns, but they are fictitious states and do not represent this persistent dynamics. \\
Secondly, if we assume that the nature of the process is not continuous and repeated values do represent the configuration of the time series, extending the alphabet could be under consideration, but noticing that the assumptions made by Bandt \& Pompe are not complied.\\
Furthermore, as stated in section \ref{OP}, $X_m(t)=(x_t,x_{t+1},\dots,x_{t+m-1})$ can be mapped to an element $\pi_i$ in $S_m$. This mapping should be defined in a way that preserves the desired relation  between the elements $x_t$ in $X_m(t)$; and all $t \in T$ that shares this pattern has to mapped to the same element of $\pi_i \in S_m$. But is also highly desirable that every $\pi_i \in S_m$ represents an unique desired structure in $X_m(t)$, leading to a \textit{"bijective"} mapping between pattern structures and symbols. When the alphabet is extended using the order (i.e \textit{order based alphabet}) as implemented in \citet{bian2012modified} this desirable condition does not comply for every embedding dimension $m$. Even more, as $m$ increases the number of different structures of the patterns $X_t(m)$ that maps onto the same symbol increases. In fact, this number is quantified in table \ref{table:cantsymbols} by the difference between the amount of symbols in the rank extended alphabet and the order extended alphabet. 
For example for $m=4$ there are $73$ symbols in the \textit{chronological extended alphabet} and $75$ symbols in the \textit{rank extended alphabet}. This is because in the \textit{chronological extended alphabet} $X_4(t_1)=(1,4,1,4)$ and $X_4(t_2)=(1,4,4,1)$ are both mapped to $\pi_i=(1122)$ and  $X_4(t_3)=(4,1,4,1)$ and $X_4(t_4)=(4,1,1,4)$ are both mapped to $\pi_i=(2211)$. It can be easily seen in table \ref{table:biject}, that $X_4(t_1)$ and $X_4(t_2)$ have a completely different relation between their elements, but yet they are mapped to the same symbol, and the same occurs with $X_4(t_3)$ and $X_4(t_4)$.
In the \textit{rank extended alphabet} $X_4(t_1),X_4(t_2),X_4(t_3),X_4(t_4)$ are mapped to four different symbols ($(1313),(1331),(3131),(3113)$) at least preserving the distinct structures of the patterns. 

\begin{table}[]
\centering

\begin{tabular}{|m{20mm}|c|c|c|}
\hline
Pattern                                & \begin{tabular}[c]{@{}c@{}}Chronological\\ Extended\\ Alphabet\end{tabular} & \begin{tabular}[c]{@{}c@{}}Rank \\ Extended\\ Alphabet\end{tabular} &$X_4(t)$\\ \hline
\includegraphics[width=20mm]{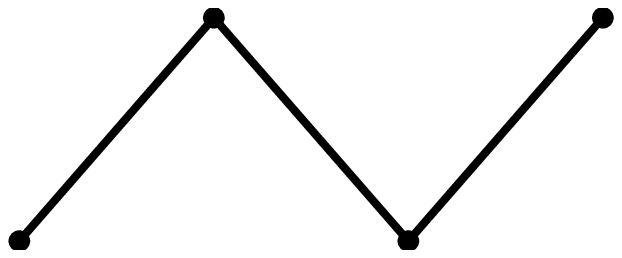} & 1122                                                                    & 1313                                                                   & $X_4(t_1)$\\ \cline{1-1} \cline{3-3} 
\includegraphics[width=20mm]{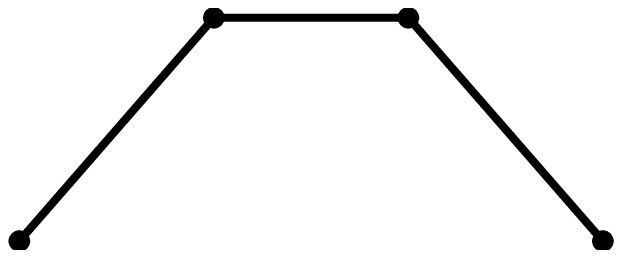} &                                                                         & 1331                                                                   &$X_4(t_2)$ \\ \hline
\includegraphics[width=20mm]{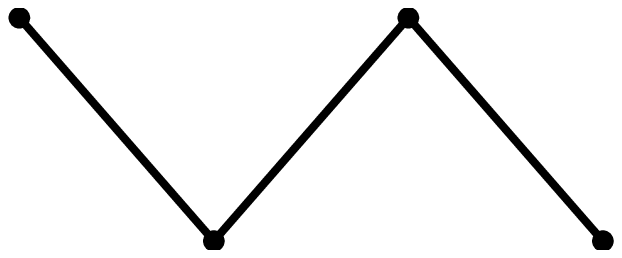} & 2211                                                                    & 3131                                                                   &$X_4(t_3)$ \\ \cline{1-1} \cline{3-3} 
\includegraphics[width=20mm]{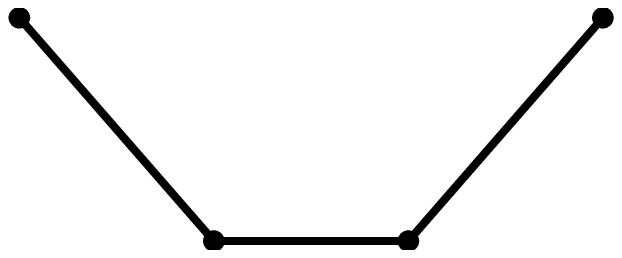} &                                                                         & 3113                                                                   &$X_4(t_4)$ \\ \hline
\end{tabular}\label{table:biject}\caption{Chronological extended alphabet proposed in \citet{bian2012modified} produces a mapping that represents patterns with different relation in the structure onto the same symbol. This is an example for $m=4$ where two structures are wrongly  mapped. As $m$ increases the number of different structures of the patterns $X_t(m)$ that maps onto the same symbol increases. In fact, this number is quantified in table \ref{table:cantsymbols} by the difference between the amount of symbols in the rank extended alphabet and the order extended alphabet.}
\end{table}

Extending the alphabet proves not to be a fine tool in order to estimate the Permutation Entropy of the original series, at least if one assume that the repeated values are due to observational problems. In the case this methodology is used we recommend not to implement it as is proposed by \citet{bian2012modified}, but with the Extended Rank Alphabet that preserves the dynamical structure of the embedded vectors $X_t(m)$.\\

When dealing with Bandt \& Pompe Permutation Entropy values with ties should be treated as missing values. Eliminating all the patterns with ties (Complete Cases methodology) is a good estimator of $H(m)$ as long as the length of the time series is much larger than the amount of deleted cases, so even though those cases are not negligible, its elimination does not affect the dynamics of the process. When this is not the case, with this methodology, the bias of $\hat{H}$ is not constant along the actual value of $H(m)$ leading to unwanted properties of that estimator.\\
Random Imputation methodology overestimates the entropy as it adds random noise to the series in order to break equalities, but it also mask forbidden patters, by inducing those missing patterns to appear, concealing this dynamical property. 
The two remaining methodologies, Time Ordered Imputation and Bayesian Imputation, prove to be good estimators of $H(m)$ in every condition, but we recommend to use the Bayesian Imputation methodology as it does not incur to artificial ordering. 

\begin{table*}[]
\centering
\caption{All possible patterns  for $m=3$ are shown in the first column, and their mapping for every methodology are presented. For Random Imputation if a pattern has a repeated value, it is mapped randomly to a suitable symbol. If the length of the series is large, the frequency of each suitable symbol is the same with probability 1. Instead, for the Bayesian imputation, the probabilities used in order to choose a suitable symbol are derived for the Complete Cases methodology, taking account of the structure of the Time Series, preserving, for example, the missing patterns. }
\label{all3patterns}
\begin{tabular}{|m{20mm}|c|c|c|c|c|c|}
\hline
PATTERN                               & \begin{tabular}[c]{@{}c@{}}Chrono.\\  Ext.\end{tabular} & \begin{tabular}[c]{@{}c@{}}Rank\\ Ext.\end{tabular} & \begin{tabular}[c]{@{}c@{}}C.\\ Cases\end{tabular} & \begin{tabular}[c]{@{}c@{}}Time \\ Ordered\end{tabular} & \begin{tabular}[c]{@{}c@{}}Random \\ Imputation\end{tabular}   & \begin{tabular}[c]{@{}c@{}}Bayesian\\ Imputation\end{tabular}         \\ \hline
                                      & $\pi_i$                                               & $\pi_i$                                             & $\pi_i$                                            & $\pi_i$                                                 & $\#$                                                           & $\#$                                                               \\ \hline
\includegraphics[width=20mm]{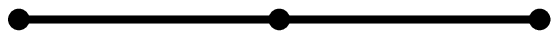} & 111                                                   & 111                                                 & X                                                  & to 123                                                  & \begin{tabular}[c]{@{}c@{}}to anyone \\ $p = 1/6$\end{tabular} & \begin{tabular}[c]{@{}c@{}}to anyone\\ $p=p^{*}(\pi_i)$\end{tabular} \\ \hline
\includegraphics[width=20mm]{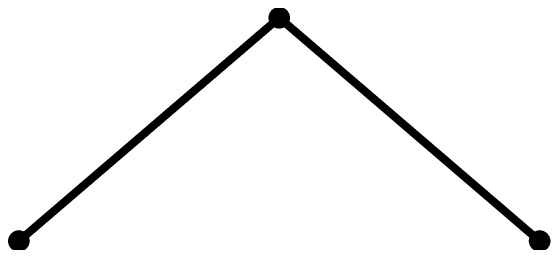} & 112                                                   & 131                                                 & X                                                  & to 132                                                  & \begin{tabular}[c]{@{}c@{}}to 132 \tiny{$p = 1/2$}\\ to 231 \tiny{$p = 1/2$} \end{tabular}       & \begin{tabular}[c]{@{}c@{}}to 132  \tiny{$p = p^*(132)$}\\ to 231 \tiny{$p = p^*(231)$}\end{tabular}            \\ \hline
\includegraphics[width=20mm]{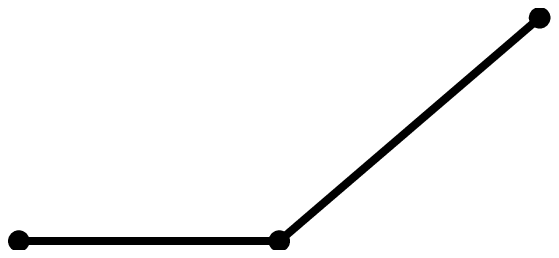} & 113                                                   & 113                                                 & X                                                  & to 123                                                  & \begin{tabular}[c]{@{}c@{}}to 123 \tiny{$p = 1/2$}\\ to 213 \tiny{$p = 1/2$}\end{tabular}        & \begin{tabular}[c]{@{}c@{}}to 123 \tiny{$p = p^*(123)$}\\ to 213 \tiny{$p = p^*(213)$}\end{tabular}            \\ \hline
\includegraphics[width=20mm]{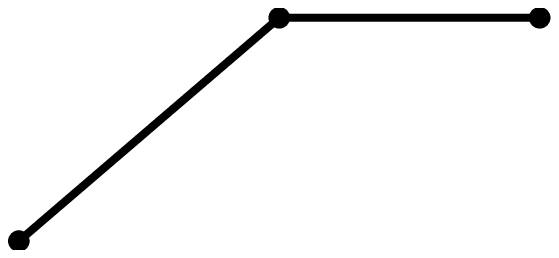} & 122                                                   & 122                                                 & X                                                  & to 123                                                  & \begin{tabular}[c]{@{}c@{}}to 123 \tiny{$p = 1/2$}\\ to 132 \tiny{$p = 1/2$}\end{tabular}        & \begin{tabular}[c]{@{}c@{}}to 123 \tiny{$p = p^*(123)$}\\ to 132 \tiny{$p = p^*(132)$}\end{tabular}            \\ \hline
\includegraphics[width=20mm]{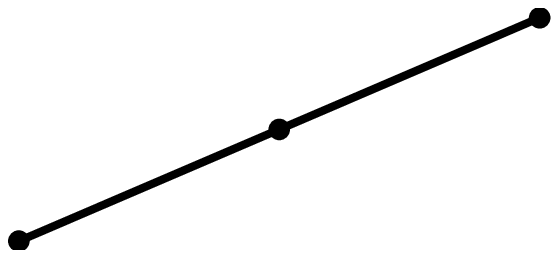} & 123                                                   & 123                                                 & 123                                                & 123                                                     & 123                                                            & 123                                                                \\ \hline
\includegraphics[width=20mm]{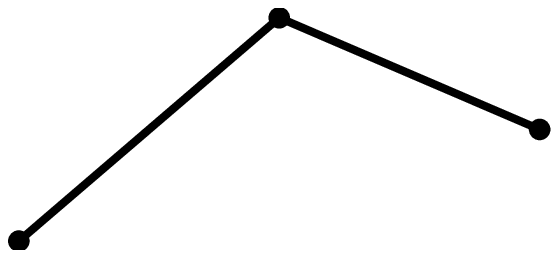} & 132                                                   & 132                                                 & 132                                                & 132                                                     & 132                                                            & 132                                                                \\ \hline
\includegraphics[width=20mm]{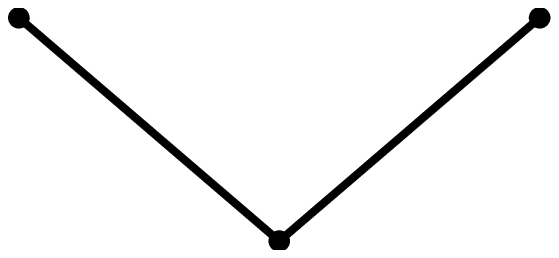} & 211                                                   & 212                                                 & X                                                  & to 213                                                  & \begin{tabular}[c]{@{}c@{}}to 213 \tiny{$p = 1/2$}\\ to 312 \tiny{$p = 1/2$}\end{tabular}        & \begin{tabular}[c]{@{}c@{}}to 213 \tiny{$p = p^*(213)$}\\ to 312 \tiny{$p = p^*(312)$}\end{tabular}            \\ \hline
\includegraphics[width=20mm]{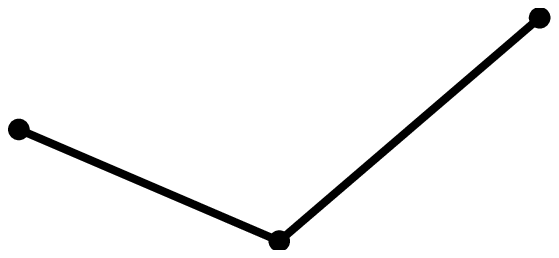} & 213                                                   & 213                                                 & 213                                                & 213                                                     & 213                                                            & 213                                                                \\ \hline
\includegraphics[width=20mm]{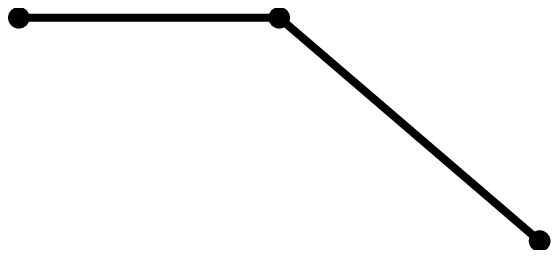} & 221                                                   & 311                                                 & X                                                  & to 312                                                  & \begin{tabular}[c]{@{}c@{}}to 312 \tiny{$p = 1/2$}\\ to 321 \tiny{$p = 1/2$}\end{tabular}        & \begin{tabular}[c]{@{}c@{}}to 312 \tiny{$p = p^*(312)$}\\ to 321 \tiny{$p = p^*(321)$}\end{tabular}            \\ \hline
\includegraphics[width=20mm]{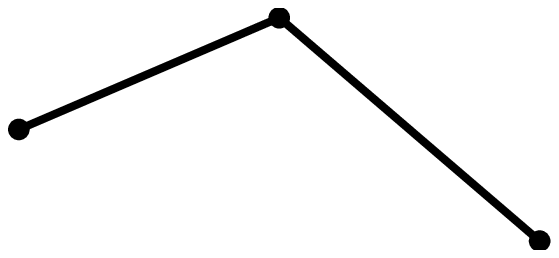} & 231                                                   & 312                                                 & 312                                                & 312                                                     & 312                                                            & 312                                                                \\ \hline
\includegraphics[width=20mm]{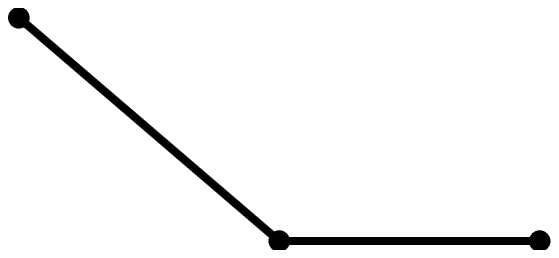} & 311                                                   & 221                                                 & X                                                  & to 231                                                  & \begin{tabular}[c]{@{}c@{}}to 231 \tiny{$p = 1/2$}\\ to 321 \tiny{$p = 1/2$}\end{tabular}        & \begin{tabular}[c]{@{}c@{}}to 231 \tiny{$p = p^*(231)$}\\ to 321 \tiny{$p = p^*(321)$}\end{tabular}            \\ \hline
\includegraphics[width=20mm]{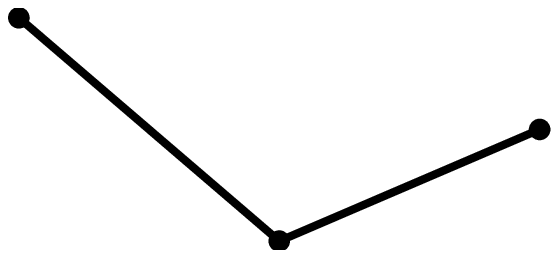} & 312                                                   & 231                                                 & 231                                                & 231                                                     & 231                                                            & 231                                                                \\ \hline
\includegraphics[width=20mm]{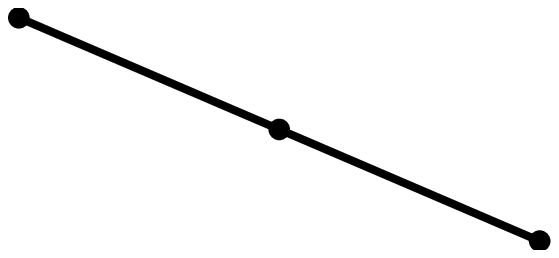} & 321                                                   & 321                                                 & 321                                                & 321                                                     & 321                                                            & 321                                                                \\ \hline
\end{tabular}
\end{table*}
\bibliography{bibliografia}

\end{document}